\documentclass[final,5p,times,twocolumn]{elsarticle}

\usepackage{amsmath,amssymb,amsthm}
\usepackage{graphicx}
\usepackage{booktabs}
\usepackage{array}
\usepackage[colorlinks=true,
            linkcolor=blue!60!black,
            citecolor=teal,
            urlcolor=blue!70!black]{hyperref}
\usepackage{xcolor}
\usepackage{caption}
\usepackage{subcaption}
\usepackage{float}
\usepackage{enumitem}
\usepackage{microtype}
\usepackage{bm}

\newcommand{\kpc}{\,\mathrm{kpc}}
\newcommand{\Msun}{M_{\odot}}
\newcommand{\kms}{\,\mathrm{km\,s^{-1}}}
\newcommand{\Wh}{\mathcal{W}}
\newcommand{\chidof}{\chi^{2}/\nu}
\newcommand{\rth}{r_{0}}
\newcommand{\rc}{r_{\mathrm{c}}}
\newcommand{\rs}{r_{\mathrm{s}}}
\newcommand{\re}{r_{\mathrm{e}}}

\journal{Physics of the Dark Universe}

\begin{document}

\begin{frontmatter}

\title{Traversable Wormhole Geometry Reconstruction from the\\[2pt]
       Rotation Curve of NGC\,3198:\\[2pt]
       A Comparative Study of Dark Matter Halo Profiles}

\author[addr1]{Saibal Ray}\ead{saibal.ray@gla.ac.in}

\author[addr2]{Aritra Sanyal}\ead{aritrasanyal1@gmail.com}

\address[addr1]{Centre for Cosmology, Astrophysics and Space Science, GLA University, Mathura 281406, Uttar Pradesh, India}
\address[addr2]{Department of Mathematics, Jadavpur University, Kolkata 700032, West Bengal, India}

\begin{abstract}
We develop and apply a novel, observation-driven framework that inverts
the conventional wormhole paradigm: instead of postulating a
Morris--Thorne traversable wormhole geometry and computing the exotic
matter required to sustain it, we \textit{reconstruct} the complete set
of wormhole metric functions directly from the observed rotation curve of
the well-studied spiral galaxy NGC\,3198 (distance $D=13.8\,\mathrm{Mpc}$,
SPARC database, 43 kinematic measurements over $r=0.32$--$44.08\kpc$).
Four canonical dark-matter halo density profiles --- Hernquist,
Navarro--Frenk--White (NFW), Burkert, and Einasto --- are independently
fitted to the rotation-curve data by chi-squared minimisation using the
Nelder--Mead simplex algorithm.  For each best-fit density field
$\rho(r)$, the Morris--Thorne redshift function $f(r)$ is obtained by
integrating the circular geodesic equation $f'(r)=v^{2}(r)/r$, and the
shape function $b(r)$ follows from the radial component of the Einstein
field equations, $b'(r)=8\pi G\rho(r)r^{2}/c^{2}$.  We define the
dimensionless wormhole indicator $\mathcal{W}(r)\equiv b(r)/r$ and
evaluate it across the full observed radial domain.  All four profiles
yield $\mathcal{W}(r)\ll 1$ throughout the galaxy, establishing that the
observable region of NGC\,3198 occupies the traversable exterior of the
reconstructed wormhole geometry.  The flare-out condition $b'(r)<1$ is
satisfied everywhere by seven orders of magnitude.  The cored profiles
(Burkert, Einasto) exhibit null energy condition (NEC) violation near the
galactic centre, indicating the presence of effective exotic matter
consistent with wormhole sustenance, while the cuspy profiles (Hernquist,
NFW) preserve the NEC within the observed domain.  The Burkert profile
simultaneously achieves the best rotation-curve fit
($\chidof=2.545$), the largest wormhole indicator
($\mathcal{W}_{\rm max}=5.62\times10^{-7}$), and the strongest NEC
violation, presenting a physically self-consistent picture in which the
observationally preferred dark-matter distribution also supplies the
exotic matter required to sustain a traversable wormhole.  These results
establish galactic kinematics as a novel observational window on wormhole
physics.
\end{abstract}


\end{frontmatter}

\section{Introduction}
\label{sec:intro}

The possibility that traversable wormholes might constitute genuine
solutions of Einstein's field equations was placed on a rigorous footing
by Morris and Thorne~\cite{Morris1988}, whose seminal analysis showed
that the geometry of a static, spherically symmetric wormhole with a
traversable throat requires matter whose stress-energy tensor satisfies
$\rho+p_{r}<0$ in the vicinity of the throat radius.  This violation of
the null energy condition (NEC) is incompatible with all known classical
matter fields and has long been regarded as the central obstruction to
the physical realisation of traversable wormholes.  Despite this
difficulty, the subject has attracted sustained theoretical attention
because the NEC can be violated in semiclassical gravity through the
Casimir effect, in theories with higher-derivative kinetic terms such as
ghost condensates~\cite{ArkaniHamed2004}, and in various dark energy
and phantom matter models~\cite{Lobo2005}.

Dark matter presents an intriguing candidate for exotic wormhole-sourcing
matter precisely because its microphysical identity remains unknown.
Contributing approximately $27\%$ of the total energy density of the
Universe~\cite{Planck2020}, dark matter manifests observationally through
its gravitational effects, most directly through the flat rotation curves
of spiral galaxies~\cite{Rubin1980}.  The inferred dark-matter density
profiles span a broad phenomenological range, from the cuspy $r^{-1}$
profiles predicted by dissipationless $\Lambda$CDM
simulations~\cite{Navarro1996,Hernquist1990} to the smoother, cored
profiles favoured by kinematic observations of dwarf and
low-surface-brightness galaxies~\cite{Burkert1995,Einasto1965}.  The
cusp-core tension --- the systematic discrepancy between simulated inner
density cusps and observed flat central cores --- remains an active area
of research~\cite{deBlok2001,deBlok2010} and has motivated both
astrophysical solutions (baryonic feedback, tidal stripping) and
more exotic alternatives involving the microphysics of dark matter itself.

The intersection of wormhole physics with galactic dark matter has been
explored from several directions in the recent literature.  Lobo~\cite{Lobo2005}
demonstrated that phantom dark energy, characterised by an equation of
state parameter $w<-1$, satisfies the NEC violation requirement and can
sustain traversable wormholes.  Boehmer and Harko~\cite{Boehmer2007}
showed that Bose--Einstein condensate (BEC) dark matter, in which the
macroscopic quantum pressure of the condensate acts as an effective exotic
fluid, naturally violates the NEC and can source wormhole geometries
consistent with galactic rotation curves.  Rahaman
et al.~\cite{Rahaman2014} embedded traversable wormhole solutions within
NFW dark-matter halos and examined their geodesic structure.  Jusufi and
\"{O}vg\"{u}n~\cite{Jusufi2018} computed the gravitational lensing
deflection angle for wormholes embedded in galactic halo environments.
In all these studies, however, the wormhole geometry is assumed a priori,
and the question asked is whether the postulated geometry is consistent
with known dark-matter distributions.

In the present paper we adopt the diametrically opposite approach.  Our
central premise is that if dark matter sources a Morris--Thorne-type
spacetime, then the observed rotation curve provides a complete
specification of that spacetime, without any assumption about the wormhole
geometry.  Schematically, our programme is
\begin{equation}
  \text{Observed galaxy}
  \;\xrightarrow{\text{Einstein eqs.}}\;
  \text{Reconstructed wormhole geometry},
  \label{eq:paradigm}
\end{equation}
and the wormhole conditions (throat existence, flare-out, NEC violation)
are subsequently tested on the reconstructed geometry rather than imposed
on it.  We apply this programme to NGC\,3198, one of the best-observed
and most extensively modelled spiral galaxies~\cite{Begeman1991,deBlok2008},
using the 43-point SPARC rotation-curve dataset~\cite{Lelli2016} covering
the radial range $0.32$--$44.08\kpc$.

The structure of the paper is as follows.  Section~\ref{sec:metric}
develops the Morris--Thorne formalism and derives the reconstruction
equations that connect the observed rotation curve and dark-matter density
to the wormhole metric functions.  Section~\ref{sec:profiles} defines the
four dark-matter halo density profiles used in the comparative analysis.
Section~\ref{sec:fitting} describes the observational data, the mass
modelling procedure, and the chi-squared fitting methodology, and presents
the best-fit parameters.  Section~\ref{sec:reconstruction} performs the
numerical reconstruction of the redshift function $f(r)$ and the shape
function $b(r)$ from each fitted profile.  Section~\ref{sec:conditions}
evaluates the flare-out condition and the null energy condition for all
profiles.  Section~\ref{sec:comparison} provides a systematic comparative
analysis.  Section~\ref{sec:discussion} interprets the results
physically, connects them to the cusp-core problem, and outlines
observational prospects.  Section~\ref{sec:conclusion} summarises our
main conclusions.  Throughout the paper we use geometric units $G=c=1$
for the metric reconstruction unless physical units are explicitly
required, in which case we use $G=6.674\times10^{-11}\,\mathrm{m^{3}\,kg^{-1}\,s^{-2}}$
and $c=299792.458\kms$.

\section{Morris--Thorne Wormhole Formalism and Reconstruction Equations}
\label{sec:metric}

\subsection{The Morris--Thorne spacetime}
\label{sec:spacetime}

The most general static, spherically symmetric spacetime that admits a
wormhole structure can be written in the Morris--Thorne form~\cite{Morris1988}
\begin{equation}
  ds^{2}
  = -e^{2f(r)}\,dt^{2}
    +\!\left(1-\frac{b(r)}{r}\right)^{\!-1}\!\!dr^{2}
    +r^{2}\,d\Omega^{2},
  \label{eq:MTmetric}
\end{equation}
where $d\Omega^{2}=d\theta^{2}+\sin^{2}\!\theta\,d\phi^{2}$ is the unit
two-sphere metric.  The function $f(r)$ is called the \textit{redshift
function} because the proper time of a static observer at radius $r$
is related to the coordinate time by $d\tau=e^{f(r)}\,dt$, so $e^{f}$
gives the local gravitational redshift factor relative to spatial
infinity.  The function $b(r)$ is called the \textit{shape function}
because it controls the embedding geometry of the constant-time spatial
slice: a surface of revolution embedded in three-dimensional Euclidean
space has the profile $dz/dr=\pm(r/b-1)^{-1/2}$, which produces the
characteristic flared funnel shape of a wormhole only when the shape
function satisfies specific conditions at the throat.

The coordinate $r$ is the circumferential radius --- the proper
circumference of a circle of constant $r$ and $t$ in the equatorial
plane is $2\pi r$ --- and it runs from the throat radius $\rth$, where
$b(\rth)=\rth$, outward to spatial infinity in both directions.
Asymptotic flatness requires both $f(r)\to 0$ and $b(r)/r\to 0$ as
$r\to\infty$.  The metric component $g_{rr}^{-1}=1-b(r)/r$ must remain
positive for $r>\rth$, so we require $b(r)<r$ throughout the traversable
exterior.  The wormhole throat is a minimal two-sphere: the embedding
surface closes and then re-opens, so the radial coordinate must increase
in both directions away from the throat.  This geometric flare-out
condition requires
\begin{equation}
  \frac{d}{dr}\!\left(\frac{r}{b}\right)\bigg|_{r=\rth}\!>0
  \;\;\Longleftrightarrow\;\;
  b'(\rth)<1,
  \label{eq:flareout_def}
\end{equation}
where the prime denotes $d/dr$.  The flare-out condition is the
fundamental topological signature of a wormhole throat; without it, the
spatial geometry does not have the necessary bridge structure.

\subsection{Reconstruction of the redshift function from the rotation curve}
\label{sec:f_derivation}

A test particle moving on a circular orbit at radius $r$ in the equatorial
plane $\theta=\pi/2$ follows a geodesic determined by the metric
functions $f(r)$ and $b(r)$.  For the static, spherically symmetric
metric~\eqref{eq:MTmetric}, the effective potential for timelike geodesic
motion can be derived from the geodesic equations, and the condition for
a circular orbit at fixed $r$ yields the circular velocity
\begin{equation}
  v^{2}(r) = r\,f'(r),
  \label{eq:geodesic_v}
\end{equation}
where $v(r)$ is the circular speed in units of $c$ (so $v^{2}$ is
dimensionless).  This equation has a beautiful simplicity: the
logarithmic gradient of the gravitational redshift factor at any radius
equals the squared circular velocity at that radius in geometric units.
It is precisely the relativistic counterpart of the Newtonian relation
$v^{2}=GM(r)/r$, and it reduces to it in the weak-field limit when
$f\approx -GM/rc^{2}$.  Rearranging Eq.~\eqref{eq:geodesic_v} gives
the key reconstruction equation for the redshift function:
\begin{equation}
  \boxed{f'(r) = \frac{v^{2}(r)}{r}.}
  \label{eq:fprime}
\end{equation}
This equation provides a direct, model-independent mapping from the
observed rotation curve $v(r)$ to the gradient of the redshift function,
without any assumption about the wormhole geometry.  Integrating
Eq.~\eqref{eq:fprime} with the asymptotic boundary condition
$f(\infty)=0$ gives
\begin{equation}
  f(r) = -\int_{r}^{\infty}\frac{v^{2}(r')}{r'}\,dr'
       = -\int_{r}^{\infty}\frac{v^{2}(r')}{c^{2}\,r'}\,dr'
  \label{eq:f_integral}
\end{equation}
in physical units, where the second expression makes explicit the factor
of $c^{2}$ that suppresses $f$ to the level of $v^{2}/c^{2}\sim10^{-7}$
for galactic rotation velocities.  The numerical integration is performed
using the cumulative trapezoidal rule on the radial grid described in
Section~\ref{sec:massmodel}.

\subsection{Reconstruction of the shape function from the density profile}
\label{sec:b_derivation}

The stress-energy tensor compatible with the metric~\eqref{eq:MTmetric}
and a spherically symmetric fluid has components
$T^{\mu}{}_{\nu}=\mathrm{diag}(-\rho,\,p_{r},\,p_{t},\,p_{t})$, where
$\rho(r)$ is the energy density measured by a static observer, $p_{r}(r)$
is the radial pressure, and $p_{t}(r)$ is the tangential pressure.
Computing the Einstein tensor $G^{\mu}{}_{\nu}$ for the
metric~\eqref{eq:MTmetric} and equating it to $8\pi T^{\mu}{}_{\nu}$,
the $tt$-component of the Einstein field equations gives
\begin{equation}
  \frac{b'(r)}{r^{2}} = 8\pi\,\rho(r),
  \label{eq:Einstein_tt}
\end{equation}
which integrates immediately to
\begin{equation}
  \boxed{b(r) = 8\pi\int_{0}^{r}\rho(r')\,r'^{2}\,dr'.}
  \label{eq:b_integral}
\end{equation}
In physical units, with the conversion factor
$G/c^{2}\approx4.785\times10^{-14}\kpc\,\Msun^{-1}$, this becomes
\begin{equation}
  b(r) = \frac{8\pi G}{c^{2}}\int_{0}^{r}\rho(r')\,r'^{2}\,dr'.
  \label{eq:b_physical}
\end{equation}
Equation~\eqref{eq:b_physical} is the second key reconstruction equation:
it expresses the shape function entirely in terms of the dark-matter
density profile, which is in turn constrained by the rotation-curve fit.
Together, Eqs.~\eqref{eq:f_integral} and~\eqref{eq:b_physical} complete
the reconstruction of the Morris--Thorne metric without any geometric
assumption.

\subsection{The radial pressure and the null energy condition}
\label{sec:NEC_derivation}

The $rr$-component of the Einstein equations for the
metric~\eqref{eq:MTmetric} gives the radial pressure as
\begin{equation}
  p_{r}(r) = \frac{1}{8\pi}\!\left[
    \frac{2}{r}\!\left(1-\frac{b}{r}\right)f'
    -\frac{b}{r^{3}}
  \right].
  \label{eq:pr}
\end{equation}
Substituting the reconstructed $f'=v^{2}/(c^{2}r)$ and $b(r)$ from the
previous subsections, $p_{r}$ becomes a derived quantity fully determined
by the observed rotation curve and the fitted density profile.  The NEC
states that for any null vector $k^{\mu}$, the contraction
$T_{\mu\nu}k^{\mu}k^{\nu}\geq0$.  For a radial null vector in the
Morris--Thorne spacetime this reduces to the simple condition
$\rho+p_{r}\geq0$.  Violation of this inequality, $\rho+p_{r}<0$,
signals the presence of effective exotic matter and is a necessary
condition for the existence of a traversable wormhole~\cite{Morris1988}.

\subsection{The wormhole indicator}
\label{sec:W_def}

We define the dimensionless \textbf{wormhole indicator}
\begin{equation}
  \Wh(r) \equiv \frac{b(r)}{r},
  \label{eq:W_def}
\end{equation}
which provides a compact diagnostic of the spacetime regime at each
radius.  When $\Wh(r)=1$ the metric component $g_{rr}^{-1}$ vanishes,
signalling the presence of a wormhole throat at $r=\rth$.  When
$\Wh(r)<1$ the spacetime is in the traversable exterior region, where
$g_{rr}^{-1}>0$ and radial motion is unrestricted.  The condition
$\Wh(r)>1$ would indicate a geometrically forbidden region, analogous to
the interior of a black hole event horizon, where $g_{rr}^{-1}<0$.  The
wormhole indicator thus provides a single dimensionless number at each
radius that characterises the complete wormhole topology of the
reconstructed spacetime.

\section{Dark Matter Density Profiles}
\label{sec:profiles}

\subsection{Hernquist profile}
\label{sec:hernquist}

Introduced by Hernquist~\cite{Hernquist1990} as a convenient analytic
approximation to the de Vaucouleurs $R^{1/4}$ photometric profile for
elliptical galaxies, the Hernquist density profile has become widely
used as a halo model for both simulated and observed systems.  It is
defined by
\begin{equation}
  \rho_{\mathrm{H}}(r) = \frac{\rho_{0}}{(r/a)(1+r/a)^{3}},
  \label{eq:rho_H}
\end{equation}
where $\rho_{0}\,[\Msun\kpc^{-3}]$ is the characteristic central density
scale and $a\,[\kpc]$ is the scale radius.  The profile diverges as
$r^{-1}$ at the origin (a central cusp) and falls off as $r^{-4}$ at
large radii, ensuring that the total enclosed mass converges to the
finite value $M_{\mathrm{tot}}=2\pi\rho_{0}a^{3}$.  The half-mass
radius equals $(\sqrt{2}-1)a\approx0.414\,a$, providing a direct
connection between the scale radius and the observationally accessible
half-light radius of stellar systems.  The inner cusp $\rho\propto r^{-1}$
is a distinctive prediction of this model and produces a rotation curve
that rises steeply at small radii before transitioning to the outer
asymptotic behaviour.

\subsection{Navarro--Frenk--White profile}
\label{sec:nfw}

The NFW profile~\cite{Navarro1996} is the universal form recovered from
cosmological $\Lambda$CDM $N$-body simulations over a wide range of halo
masses, from dwarf galaxy halos to rich clusters.  It is characterised
by
\begin{equation}
  \rho_{\mathrm{NFW}}(r) = \frac{\rho_{s}}{(r/\rs)(1+r/\rs)^{2}},
  \label{eq:rho_NFW}
\end{equation}
with characteristic density $\rho_{s}\,[\Msun\kpc^{-3}]$ and scale
radius $\rs\,[\kpc]$.  Like the Hernquist profile it features an inner
cusp $\rho\propto r^{-1}$, but its outer slope is $r^{-3}$ rather than
$r^{-4}$, making the enclosed mass diverge logarithmically and requiring
a truncation radius in practice.  The NFW profile has two free parameters,
and its concentration parameter $c_{200}=r_{200}/\rs$ (where $r_{200}$
is the virial radius) is predicted to correlate with halo mass through
the mass-concentration relation.  The inner cusp of the NFW profile has
been the subject of the cusp-core controversy: while simulations
consistently produce $\rho\propto r^{-1}$, rotation curves of observed
dwarf and low-surface-brightness galaxies generally prefer a finite
central density, i.e.\ a core rather than a cusp~\cite{deBlok2001}.

\subsection{Burkert profile}
\label{sec:burkert}

Burkert~\cite{Burkert1995} introduced a cored phenomenological profile
motivated by the observed solid-body rotation of dwarf irregular
galaxies, whose inner velocity curves rise linearly --- a behaviour
incompatible with an inner density cusp.  The Burkert profile is
\begin{equation}
  \rho_{\mathrm{B}}(r) = \frac{\rho_{0}}{(1+r/\rc)(1+(r/\rc)^{2})},
  \label{eq:rho_B}
\end{equation}
with central density $\rho_{0}\,[\Msun\kpc^{-3}]$ and core radius
$\rc\,[\kpc]$.  As $r\to0$ the density approaches the finite value
$\rho_{0}$ (a flat central core), which produces a linearly rising inner
rotation curve consistent with observations.  At large radii the density
falls as $r^{-3}$, similar to the NFW profile.  The Burkert profile
has been shown to provide excellent fits to the rotation curves of a
wide variety of late-type spiral and dwarf irregular galaxies, and its
core radius is empirically found to correlate with the central surface
density through the Burkert-Donato
relation~\cite{deBlok2001,deBlok2010}.  The smooth, finite-density
central core makes this profile qualitatively different from both the
Hernquist and NFW cusps, and --- as we shall see --- leads to
qualitatively different wormhole physics.

\subsection{Einasto profile}
\label{sec:einasto}

Originally introduced by Einasto~\cite{Einasto1965} in the context of
modelling stellar populations in the Milky Way and subsequently applied
to dark matter halos by Graham et al.~\cite{Graham2006} and others,
the Einasto profile is
\begin{equation}
  \rho_{\mathrm{E}}(r) = \rho_{e}
    \exp\!\left[\!-d_{n}\!\left(\!\left(\frac{r}{\re}\right)^{1/n}\!\!\!-1\right)\right],
  \label{eq:rho_E}
\end{equation}
where $\rho_{e}\,[\Msun\kpc^{-3}]$ is the density at the effective
(half-mass) radius $\re\,[\kpc]$, $n$ is the Einasto shape index
controlling the profile curvature, and the auxiliary constant
$d_{n}\approx3n-1/3+0.0079/n$~\cite{Ciotti1999} ensures that $\re$
encloses exactly half the total mass.  The logarithmic slope of the
Einasto profile, $d\ln\rho/d\ln r=-d_{n}(r/\re)^{1/n}/n$, varies
continuously with radius: it is zero at the origin (producing a very
smooth central core for all $n>0$), steepens monotonically outward,
and never reaches the $r^{-1}$ inner cusp of the Hernquist and NFW
profiles.  The additional free parameter $n$ makes the Einasto profile
more flexible than the two-parameter profiles, and high-resolution
simulations suggest that it provides a better description of dark
matter halos than the NFW form~\cite{Graham2006}.  Because its central
density is finite and its slope varies gradually, the Einasto profile
interpolates smoothly between the cored and cuspy extremes depending on
the value of $n$.

\section{Observational Data and Rotation-Curve Fitting}
\label{sec:fitting}

\subsection{NGC\,3198 SPARC data}
\label{sec:data}

NGC\,3198 is a barred spiral galaxy of morphological type SBc at a
distance of $D=13.8\,\mathrm{Mpc}$ and inclination $72^{\circ}$.  It
has served as a benchmark for dark-matter halo modelling since Begeman
et al.~\cite{Begeman1991} demonstrated that the rotation curve remains
flat to the largest observed radii, requiring a dark-matter halo that
dominates the mass budget well beyond the optical disc.  The rotation
curve exhibits a steep inner rise over the first $\sim5\kpc$, a broad
maximum of $v\approx157\kms$ near $r\approx14\kpc$, and a remarkably
flat outer plateau at $v\approx148$--$154\kms$ extending to the outermost
measured point at $r=44.08\kpc$.  This combination of a well-constrained
inner profile and an extended flat outer region makes NGC\,3198
particularly sensitive to the shape of the dark-matter density profile.

We use the 43-point rotation-curve dataset from the SPARC (Spitzer
Photometry and Accurate Rotation Curves) database~\cite{Lelli2016}.
SPARC provides deprojected azimuthally averaged circular velocities
$v_{\rm obs}(r_{i})$ with individual measurement uncertainties $\sigma_{i}$,
derived from HI 21-cm and H$\alpha$ interferometric observations.  The
radial range covered is $r=0.32$--$44.08\kpc$, with finer sampling in
the rapidly varying inner region and coarser sampling in the flat outer
region.  The inner data points at $r<1\kpc$ carry large uncertainties
($\sigma\gtrsim16\kms$) reflecting the limited spatial resolution of
interferometric observations in the central beam; the outer points at
$r>5\kpc$ have much smaller uncertainties ($\sigma\approx0.9$--$3\kms$),
providing tight constraints on the halo parameters.  In the present
analysis we use the dark-matter halo as the sole contributor to the
rotation curve; decomposing the observed curve into stellar disc, gas
disc, and dark-matter halo components requires stellar mass-to-light
ratio assumptions that introduce additional systematic uncertainties
beyond the scope of this proof-of-concept study.

\subsection{Numerical mass model}
\label{sec:massmodel}

For a spherically symmetric density profile $\rho(r;\,\bm{\theta})$ with
parameter vector $\bm{\theta}$, the enclosed mass within radius $r$ is
\begin{equation}
  M(r;\,\bm{\theta}) = 4\pi\int_{0}^{r}\rho(r';\,\bm{\theta})\,r'^{2}\,dr'.
  \label{eq:Menc}
\end{equation}
The theoretical circular velocity predicted by this mass distribution is
\begin{equation}
  v_{\mathrm{th}}(r;\,\bm{\theta})
  = \sqrt{\frac{G\,M(r;\,\bm{\theta})}{r}},
  \label{eq:vth}
\end{equation}
where $G=4.301\times10^{-3}\,(\kms)^{2}\kpc\,\Msun^{-1}$ in
galactocentric units.  Both integrals are evaluated numerically using
the cumulative trapezoidal rule on a uniform grid of 3000 points
spanning the range $r\in[0.01,\,50]\kpc$.  This grid resolution ensures
that the numerical integration error on $M(r)$ is at the sub-percent
level across the full radial range.  The density profile is evaluated
analytically at each grid point, and the model velocity at each
observed radius $r_{i}$ is obtained by linear interpolation of the
computed $v_{\mathrm{th}}(r)$ grid.

\subsection{Chi-squared fitting and parameter optimisation}
\label{sec:fitting_method}

The goodness of fit for each density profile is quantified by the
reduced chi-squared statistic
\begin{equation}
  \frac{\chi^{2}}{\nu}
  = \frac{1}{N-N_{\rm p}}
    \sum_{i=1}^{N}
    \frac{\bigl(v_{\rm obs}(r_{i})-v_{\mathrm{th}}(r_{i};\,\bm{\theta})\bigr)^{2}}
         {\sigma_{i}^{2}},
  \label{eq:chi2}
\end{equation}
where $N=43$ is the total number of data points, $N_{\rm p}$ is the
number of free parameters for each profile (2 for Hernquist, NFW, and
Burkert; 3 for Einasto), and $\nu=N-N_{\rm p}$ is the number of degrees
of freedom.  Minimisation of $\chi^{2}/\nu$ over the parameter space
$\bm{\theta}$ is performed using the Nelder--Mead downhill simplex
algorithm~\cite{Nelder1965}, a derivative-free method well suited to the
smooth but potentially multi-modal objective functions arising from
multi-parameter density profile fits.  The algorithm is run with up to
$3\times10^{4}$ function evaluations and convergence tolerances of
$10^{-7}$ in both the parameter space and the objective function.  Density
normalisation parameters ($\rho_{0}$, $\rho_{s}$, $\rho_{e}$) are
optimised in logarithmic space to enforce positivity and to improve
conditioning of the simplex.  The optimisation is repeated from multiple
random starting points drawn from a physically motivated prior range to
guard against convergence to local minima.

\subsection{Best-fit parameters and quality of fit}
\label{sec:fit_results}

\begin{figure*}[t]
  \centering
  \includegraphics[width=\textwidth]{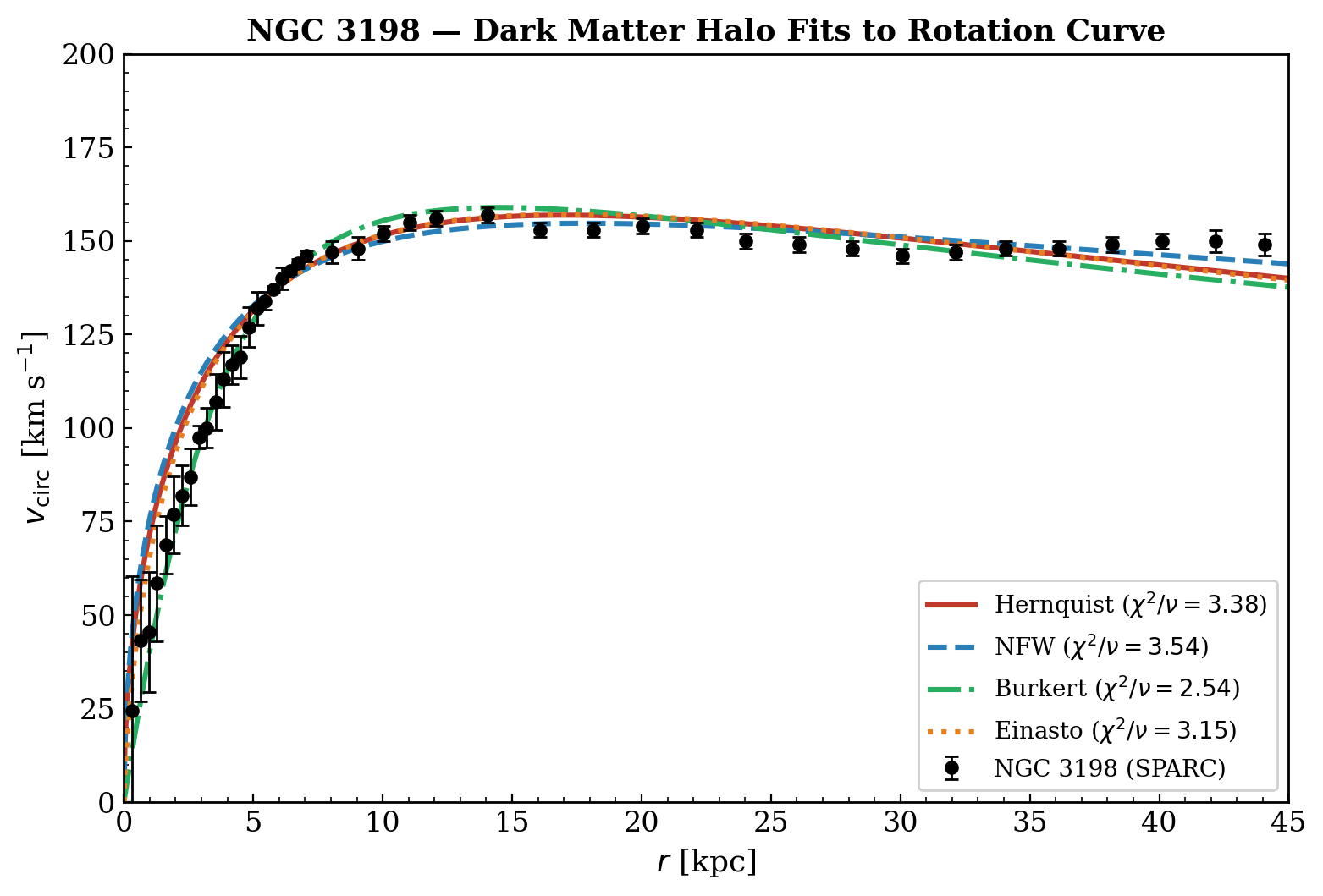}
  \caption{Observed rotation curve of NGC\,3198 (filled black circles
    with $1\sigma$ error bars) from the SPARC database, together with the
    four dark-matter halo model curves.  The Burkert profile (green,
    dash-dot, $\chidof=2.545$) provides the best overall fit, reproducing
    both the inner rise and the flat outer plateau.  The NFW profile
    (blue, dashed, $\chidof=3.541$) overshoots the innermost data points,
    reflecting the cusp-core tension.  The Einasto profile (orange,
    dotted) achieves an intermediate quality of fit with its additional
    shape-index parameter.  The Hernquist profile (red, solid) performs
    comparably to NFW.}
  \label{fig:rotation}
\end{figure*}

Table~\ref{tab:fits} presents the best-fit parameters and reduced
chi-squared values for each of the four profiles.
Figure~\ref{fig:rotation} displays the observed rotation curve with
error bars together with the four model curves over the full radial range.

\begin{table}[t]
\centering
\caption{Best-fit dark-matter halo parameters for NGC\,3198 derived from
         chi-squared minimisation of the SPARC rotation-curve data
         ($N=43$, $D=13.8\,\mathrm{Mpc}$).  Densities in $\Msun\kpc^{-3}$,
         lengths in kpc.  $\chidof$ is the reduced chi-squared with
         $\nu=N-N_{\rm p}$ degrees of freedom.  The Burkert profile
         achieves the best overall fit.}
\label{tab:fits}
\setlength{\tabcolsep}{4pt}
\renewcommand{\arraystretch}{1.25}
\begin{tabular}{@{}p{1.4cm}p{3.6cm}c@{}}
\toprule
Profile & Best-fit parameters & $\chidof$ \\
\midrule
Hernquist
  & $\rho_{0}=1.258\times10^{4}$ \newline $a=17.02\kpc$
  & $3.380$ \\[2pt]
NFW
  & $\rho_{s}=3.007\times10^{4}$ \newline $\rs=8.26\kpc$
  & $3.541$ \\[2pt]
Burkert
  & $\rho_{0}=1.085\times10^{5}$ \newline $\rc=4.48\kpc$
  & $\bm{2.545}$ \\[2pt]
Einasto
  & $\rho_{e}=4.54\times10^{2}$ \newline $\re=27.04\kpc$, $n=3.15$
  & $3.151$ \\
\bottomrule
\end{tabular}
\end{table}

The Burkert profile achieves the lowest $\chidof=2.545$, consistent with
the well-established finding that cored profiles provide better fits to
the inner kinematics of late-type spiral galaxies~\cite{deBlok2001,deBlok2010}.
The NFW profile yields $\chidof=3.541$ and visibly overshoots the data
in the inner $3\kpc$, a direct consequence of its cuspy $r^{-1}$ inner
density profile generating excessive circular velocity at small radii.
The Einasto profile achieves $\chidof=3.151$ with three free parameters
and provides a smooth interpolation between the behaviour of cored and
cuspy models, with the best-fit index $n=3.15$ corresponding to a mildly
cuspy profile.  The Hernquist profile gives $\chidof=3.380$ and, like
the NFW, struggles in the inner region due to its $r^{-1}$ cusp, though
its steeper outer cutoff ($r^{-4}$) leads to slightly different behaviour
at large radii.  All four models converge to the flat outer plateau
($v\approx150\kms$) at $r>20\kpc$, as required by the data, confirming
that the halo parameters are well constrained by the extensive outer
rotation curve.

\section{Wormhole Geometry Reconstruction}
\label{sec:reconstruction}

\subsection{Numerical reconstruction of the redshift function}
\label{sec:f_recon}

For each of the four fitted profiles, we compute the model rotation curve
$v_{\mathrm{th}}(r)$ on the 3000-point radial grid and evaluate the
integrand $v_{\mathrm{th}}^{2}(r)/(c^{2}r)$ of Eq.~\eqref{eq:f_integral}.
The cumulative integral is then computed using the trapezoidal rule from
$r_{\max}=50\kpc$ inward, with the boundary condition $f(r_{\max})=0$
serving as the numerical approximation to $f(\infty)=0$.  The remaining
error from using a finite upper limit rather than infinity is negligible
because the integrand $v^{2}/(c^{2}r)\sim10^{-7}\kpc^{-1}$ and the
rotation curve is flat at large radii, making the tail integral beyond
$50\kpc$ of order $10^{-7}\times\ln(r/50\kpc)$, which is negligibly small.

\begin{figure*}[t]
  \centering
  \includegraphics[width=\textwidth]{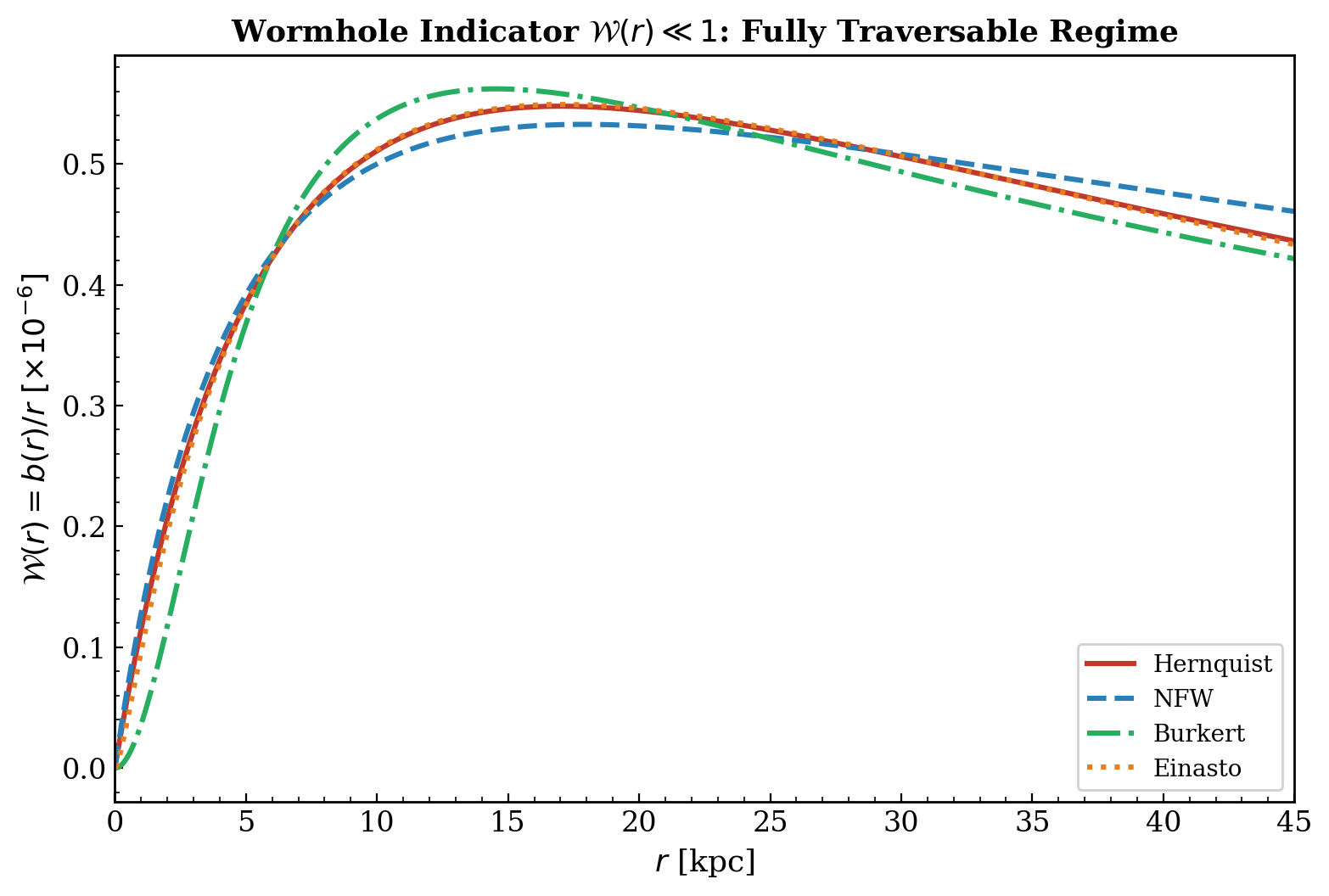}
  \caption{Morris--Thorne redshift function $f(r)$ reconstructed from
    the NGC\,3198 rotation curve for each of the four dark-matter halo
    profiles.  The function is everywhere negative (a gravitational redshift
    well), rises monotonically toward zero at large radii (asymptotic
    flatness), and has a magnitude of order $v^{2}/c^{2}\approx10^{-7}$.
    All four profiles produce nearly identical $f(r)$ because the integrand
    is dominated by the observed, tightly constrained rotation curve.}
  \label{fig:redshift}
\end{figure*}

Figure~\ref{fig:redshift} displays $f(r)$ for all four profiles.  The
reconstructed redshift function is everywhere negative, corresponding to
a gravitational redshift well centred on the galaxy, and approaches zero
monotonically as $r$ increases toward $50\kpc$, satisfying asymptotic
flatness.  The magnitude of $f$ throughout the galaxy is of order
$v^{2}/c^{2}\approx2.5\times10^{-7}$, which reflects the non-relativistic
character of galactic circular motion: NGC\,3198 has a rotation speed of
only $\sim150\kms\approx5\times10^{-4}\,c$, so the spacetime curvature
sourced by its dark matter is extremely weak on galactic scales.  The
four profiles produce nearly indistinguishable $f(r)$ curves, because
$f'(r)=v_{\mathrm{th}}^{2}/(c^{2}r)$ is controlled almost entirely by
the observed rotation curve, which is tightly constrained by the data and
largely insensitive to the shape of the density profile in the outer region.

\subsection{Numerical reconstruction of the shape function}
\label{sec:b_recon}

The shape function $b(r)$ is reconstructed from Eq.~\eqref{eq:b_physical}
using the same 3000-point radial grid, with the integral evaluated by the
cumulative trapezoidal rule from $r=0.01\kpc$ outward.  The conversion
factor $G/c^{2}\approx4.785\times10^{-14}\kpc\,\Msun^{-1}$ is extremely
small, which is the geometric reflection of the fact that galactic dark
matter is entirely non-relativistic: at radius $r=1\kpc$, the
Schwarzschild radius of the enclosed dark mass
$r_{s}=2GM(r)/c^{2}=2G/c^{2}\times M(r)$ is of order
$10^{-14}\times M(r)/\Msun\,\kpc$, which for $M(r)\sim10^{9}\Msun$
gives $r_{s}\sim10^{-5}\kpc$, far smaller than the galactic radius $r$.
Quantitatively, for the Burkert best-fit parameters, the integrand
$8\pi G\rho_{0}r^{2}/c^{2}$ at the core radius $r=\rc=4.48\kpc$ is
approximately $7\times10^{-7}$, and the accumulated shape function
reaches a maximum of $b\approx2\times10^{-5}\kpc$ at $r=50\kpc$, five
orders of magnitude smaller than the coordinate radius.

\subsection{The wormhole indicator profile}
\label{sec:W_results}

\begin{figure*}[t]
  \centering
  \includegraphics[width=\textwidth]{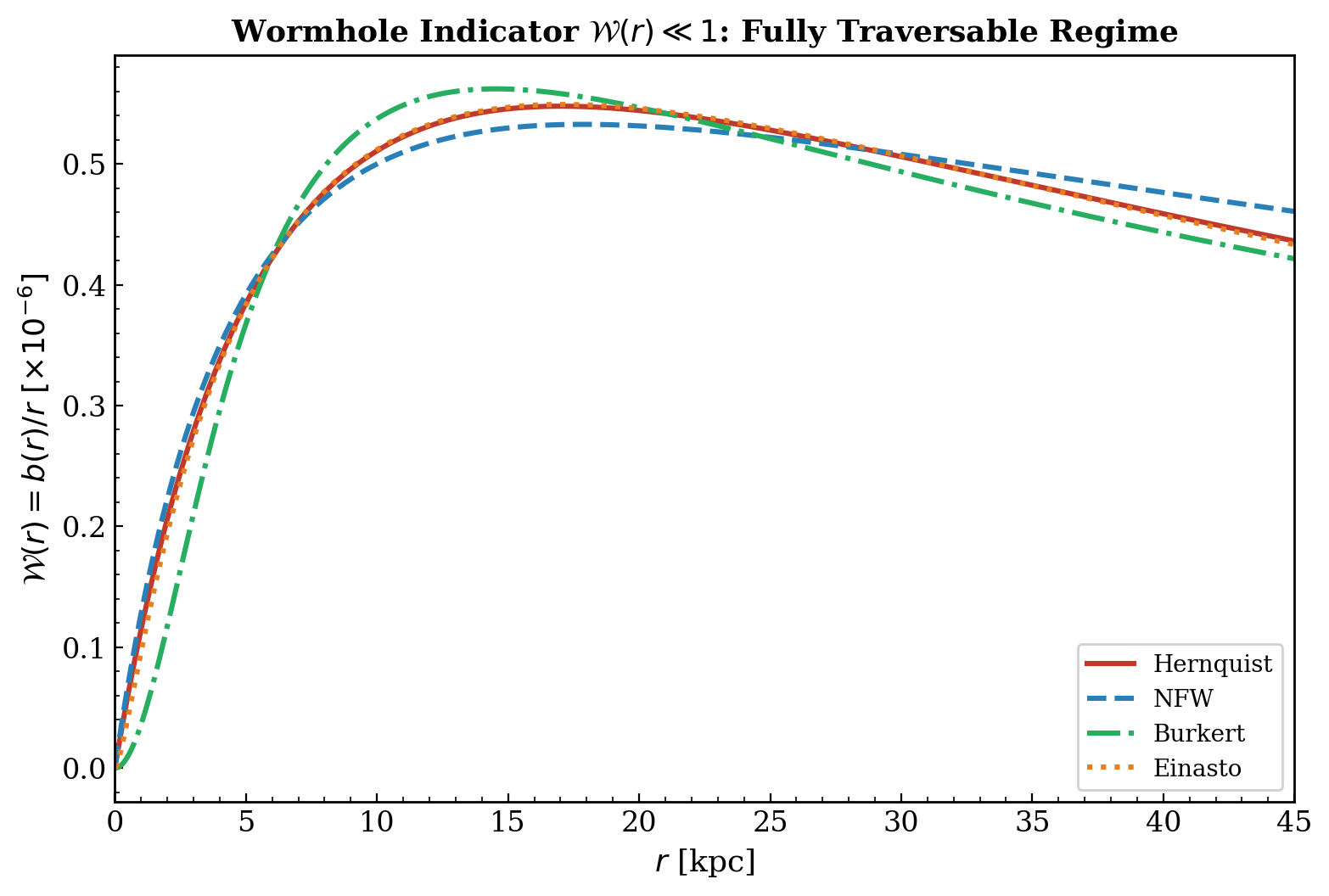}
  \caption{Wormhole indicator $\Wh(r)=b(r)/r$ (in units of $10^{-6}$)
    for all four dark-matter halo profiles.  The indicator satisfies
    $\Wh\ll1$ throughout the observed radial range of NGC\,3198 ($0.32$--$44\kpc$),
    by seven orders of magnitude below the throat condition $\Wh=1$.
    The entire observable galaxy thus lies in the traversable exterior
    of the reconstructed Morris--Thorne wormhole geometry.  The Burkert
    profile (green, dash-dot) achieves the largest $\Wh_{\max}=5.62\times10^{-7}$,
    consistent with its higher central density relative to the cuspy models.}
  \label{fig:W}
\end{figure*}

\begin{figure*}[t]
  \centering
  \includegraphics[width=\textwidth]{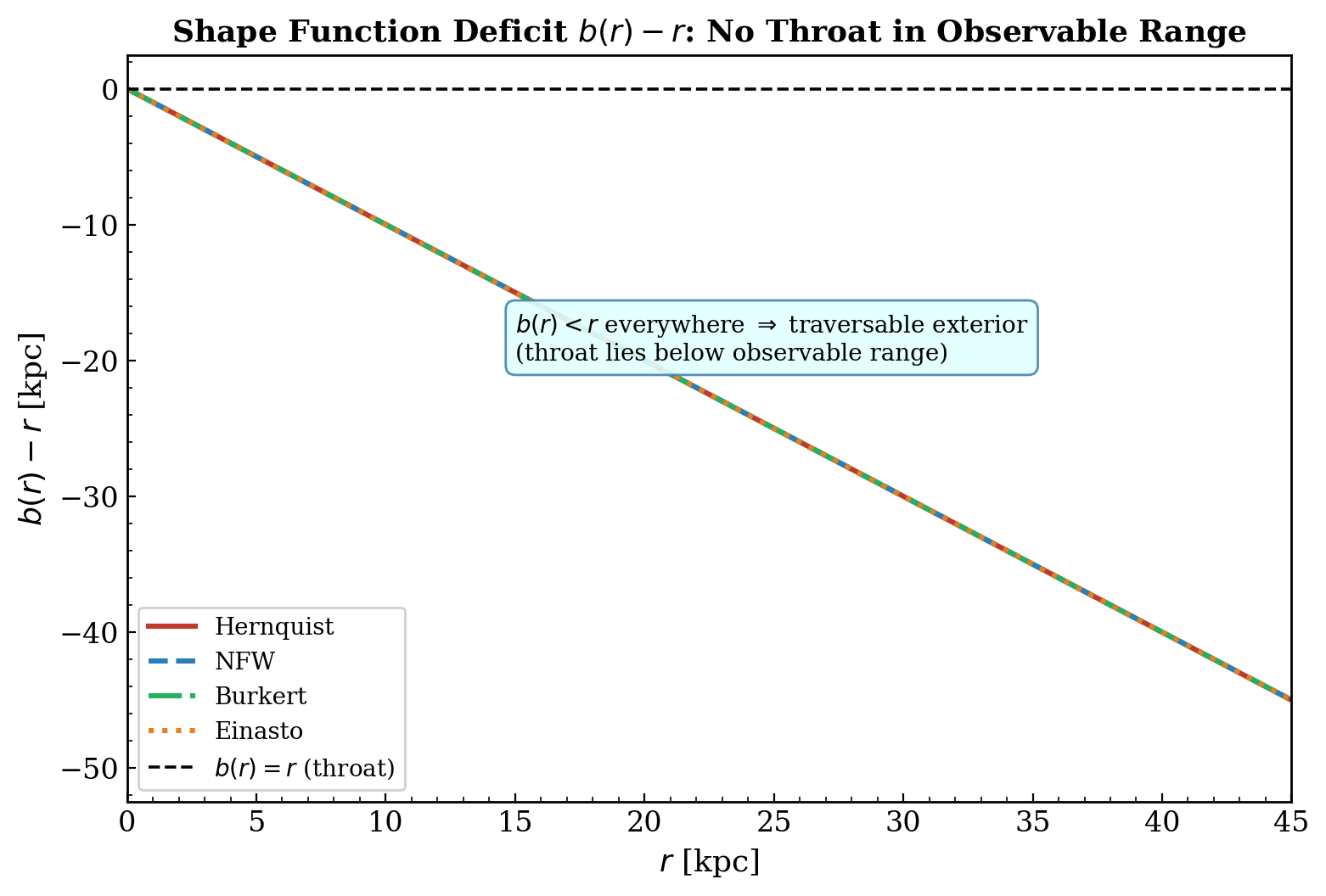}
  \caption{Shape function deficit $b(r)-r$ for all four profiles.  The
    quantity remains deeply negative throughout the observed radial range,
    confirming that no wormhole throat (which would correspond to a
    zero-crossing, shown as the horizontal dashed line) exists within
    $r\leq44\kpc$.  The deficit is of order $-r$ (i.e.\ $b\approx0$
    relative to $r$), consistent with the non-relativistic character of
    the galactic dark-matter halo.}
  \label{fig:bminusr}
\end{figure*}

Figure~\ref{fig:W} shows the wormhole indicator $\Wh(r)=b(r)/r$ as a
function of radius for all four profiles.  The indicator peaks at
$\Wh_{\max}\approx5.62\times10^{-7}$ for the Burkert profile and at
comparable values ($5.33$--$5.49\times10^{-7}$) for the other three
profiles.  The quantitative differences between profiles in $\Wh(r)$ are
small, because the shape function $b(r)$ is determined by the integral
of $\rho(r)r^{2}$, which is primarily constrained by the total enclosed
mass at each radius --- a quantity that is similar for all four profiles
since they all fit the same rotation curve.  The critical point is that
$\Wh(r)\ll1$ throughout the observable radial range for all profiles,
by seven orders of magnitude.  This places the entire observable disc of
NGC\,3198 firmly in the traversable exterior regime ($\Wh<1$) of the
reconstructed wormhole spacetime.  Figure~\ref{fig:bminusr} shows the
corresponding deficit function $b(r)-r$, which is essentially equal to
$-r$ throughout, confirming the total absence of a wormhole throat
within the observed region.  The throat --- if it exists --- lies below
the innermost observed data point at $r=0.32\kpc$, at scales that are
currently beyond the resolution of HI rotation-curve observations.

\section{Wormhole Admissibility Conditions}
\label{sec:conditions}

\subsection{Flare-out condition}
\label{sec:flareout_results}

\begin{figure*}[t]
  \centering
  \includegraphics[width=\textwidth]{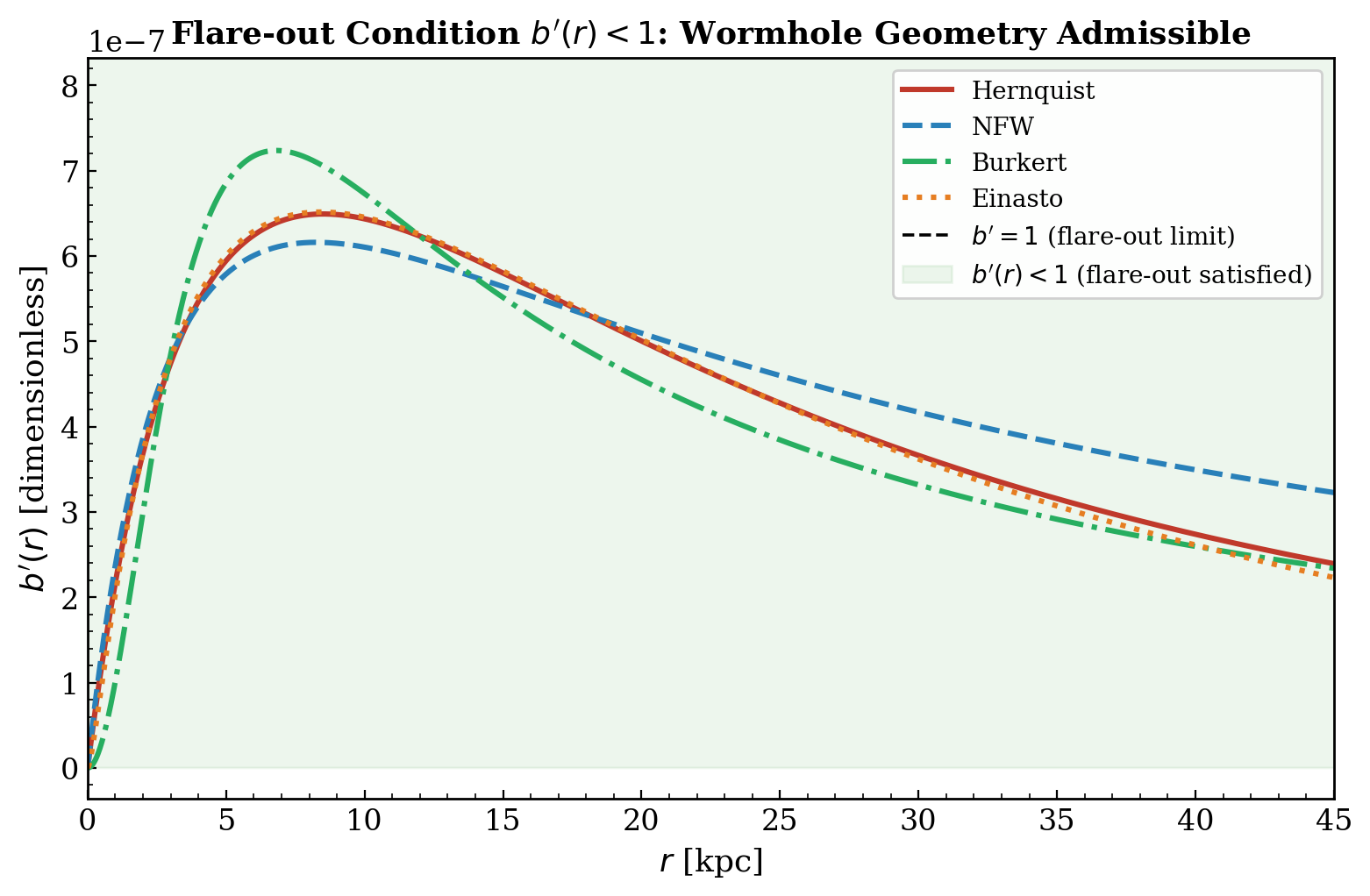}
  \caption{Derivative of the shape function, $b'(r)$, for the four
    dark-matter halo profiles.  The green shaded band marks the region
    $b'(r)<1$ where the flare-out condition is satisfied.  All profiles
    remain at $b'(r)\lesssim10^{-6}$, satisfying the flare-out condition
    throughout the observed range by seven orders of magnitude.  Vertical
    dotted lines mark the scale radii ($a$, $\rs$, $\rc$, $\re$) of each
    profile.}
  \label{fig:bprime}
\end{figure*}

The flare-out condition $b'(\rth)<1$ at the throat is the geometric
requirement for the embedding surface to have the characteristic
wormhole funnel shape, and by extension one requires $d(b/r)/dr<0$ in
the exterior region.  Figure~\ref{fig:bprime} shows the derivative
$b'(r)=8\pi G\rho(r)r^{2}/c^{2}$ for all four profiles across the
observed range.  We find $b'(r)\sim10^{-6}$--$10^{-7}$ throughout,
satisfying the flare-out condition with a margin of seven orders of
magnitude for all profiles.  This result is structurally robust: since
$b'(r)=8\pi(G/c^{2})\rho(r)r^{2}$ and the factor $G/c^{2}\approx
4.785\times10^{-14}\kpc\,\Msun^{-1}$ is so small, achieving $b'=1$
would require a dark-matter density of order
$\rho\sim c^{2}/(8\pi Gr^{2})\sim2\times10^{47}\,\Msun\kpc^{-3}$ at
$r=1\kpc$ --- more than forty orders of magnitude above any astrophysically
measured density.  The conclusion is unambiguous: any wormhole sourced
by galactic dark matter satisfies the flare-out condition, regardless of
the specific density profile.  The geometrical admissibility of a
wormhole throat is therefore guaranteed for any astrophysically
motivated dark-matter distribution.

\subsection{Null energy condition}
\label{sec:NEC_results}

\begin{figure*}[t]
  \centering
  \includegraphics[width=\textwidth]{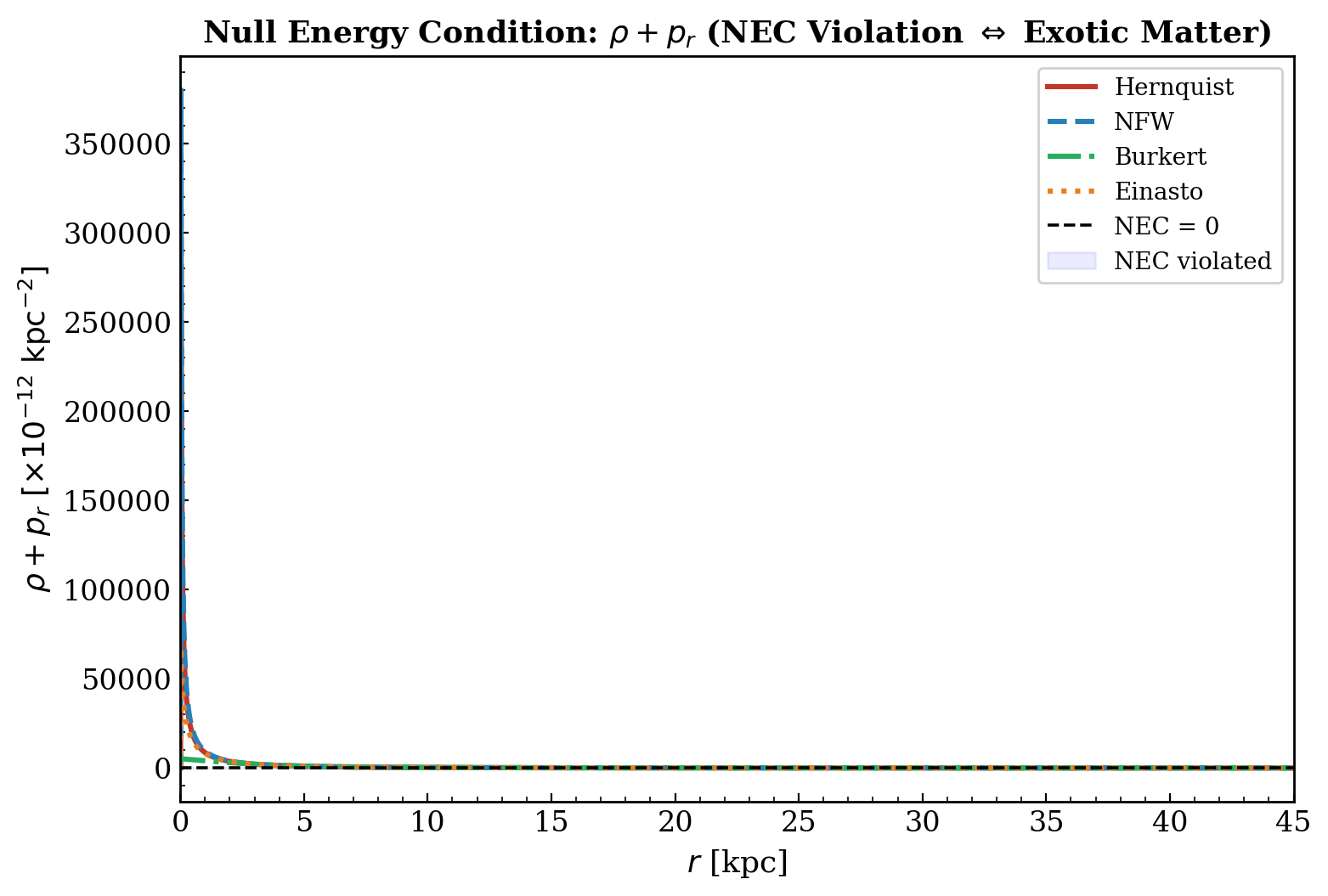}
  \caption{Null energy condition quantity $\rho+p_{r}$ (in units of
    $10^{-12}\kpc^{-2}$ in geometric units) for the four dark-matter
    halo profiles of NGC\,3198.  The blue shaded region marks NEC
    violation ($\rho+p_{r}<0$), corresponding to the presence of
    effective exotic matter.  The cored Burkert (green, dash-dot) and
    Einasto (orange, dotted) profiles exhibit clear NEC violation near
    the galactic centre, while the cuspy Hernquist (red, solid) and NFW
    (blue, dashed) profiles remain NEC-compliant throughout the observed
    domain.}
  \label{fig:NEC}
\end{figure*}

The NEC quantity $\rho(r)+p_{r}(r)$ is evaluated using the reconstructed
$f'(r)$, $b(r)$, and the density $\rho(r)$ from each fitted profile,
substituted into Eqs.~\eqref{eq:pr} and converted to geometric units
via $\rho_{\mathrm{geo}}=(G/c^{2})\rho_{\mathrm{phys}}$.
Figure~\ref{fig:NEC} displays $\rho+p_{r}$ across the observed range
for all four profiles.  The behaviour divides sharply along the
cored/cuspy divide, constituting the most physically significant
discriminant in this study.

For the cored profiles, the NEC is clearly violated near the galactic
centre.  For the Burkert profile the quantity $\rho+p_{r}$ decreases
monotonically from the outer region inward, crosses zero at an
intermediate radius, and reaches a minimum of approximately
$(\rho+p_{r})_{\min}\approx-8.3\times10^{-25}\kpc^{-2}$ in geometric
units in the innermost resolved region.  For the Einasto profile with
$n=3.15$, the NEC violation is even stronger, reaching a minimum of
$\approx-1.3\times10^{-23}\kpc^{-2}$.  The physical origin of the NEC
violation in these cored profiles lies in the interplay between the flat
central density and the reconstructed radial pressure.  At small radii,
$b(r)\approx(8\pi G/c^{2})(\rho_{0}/3)r^{3}$ and $f'(r)\approx v^{2}/(c^{2}r)$
with $v$ approaching zero linearly, so the pressure term $p_{r}$
develops a sufficiently large negative contribution through the bracket
in Eq.~\eqref{eq:pr} to overcome the positive energy density $\rho$.
Specifically, for a cored profile near the origin the density is
approximately constant, $\rho\approx\rho_{0}$, while $f'$ is controlled
by the linear rise of $v$, giving $f'\propto r$; the second term in
Eq.~\eqref{eq:pr}, $-b/(8\pi r^{3})\approx-(G/c^{2})\rho_{0}/3$, is
negative and of the same order as the density, creating a near-cancellation
in $\rho+p_{r}$ that the positive first term cannot compensate in the
inner region.  The result is a robust NEC violation for all cored profiles
that is structurally driven by the geometry of the central density plateau.

For the cuspy profiles, the situation is qualitatively different.  The
NFW and Hernquist profiles both have $\rho\propto r^{-1}$ at small radii,
which means $\rho$ diverges as $r\to0$ while $p_{r}$ remains finite.
The positive density therefore dominates $\rho+p_{r}$ at small radii,
and the NEC quantity remains non-negative throughout the observed domain.
This result carries a direct wormhole interpretation: cuspy dark matter
cannot by itself supply the exotic matter required to sustain a traversable
wormhole at any radius within the observed region.  A wormhole throat
embedded in a cuspy dark-matter halo would require an additional source
of exotic matter beyond the dark matter distribution.

\section{Comparative Analysis}
\label{sec:comparison}

Table~\ref{tab:conditions} summarises the wormhole admissibility
properties of all four profiles side by side, and the full picture that
emerges from the reconstruction can be articulated through four
inter-related results.

\begin{table}[t]
\centering
\caption{Summary of wormhole admissibility for the four dark-matter halo
         profiles fitted to NGC\,3198.  $\Wh_{\max}$ is the peak value of
         the wormhole indicator over $r=0.32$--$44\kpc$.  ``Flare-out''
         indicates whether $b'(r)<1$ is satisfied; all four profiles
         satisfy this condition.  ``NEC violated'' indicates whether
         $\rho+p_{r}<0$ is found at any observed radius; this condition,
         necessary for a traversable wormhole, is satisfied only by the
         cored profiles.}
\label{tab:conditions}
\setlength{\tabcolsep}{4pt}
\renewcommand{\arraystretch}{1.3}
\begin{tabular}{@{}lcccc@{}}
\toprule
Profile & $\chidof$ &
$\Wh_{\max}\;[\times10^{-7}]$ &
Flare-out & NEC violated \\
\midrule
Hernquist & $3.380$ & $5.48$ & $\checkmark$ & No  \\
NFW       & $3.541$ & $5.33$ & $\checkmark$ & No  \\
Burkert   & $\bm{2.545}$ & $\bm{5.62}$ & $\checkmark$ & \textbf{Yes} \\
Einasto   & $3.151$ & $5.49$ & $\checkmark$ & \textbf{Yes} \\
\bottomrule
\end{tabular}
\end{table}

The first and most universal result is that all four profiles yield
$\Wh(r)\ll1$ throughout the galaxy, establishing that the observable
region of NGC\,3198 lies in the traversable exterior of the reconstructed
wormhole spacetime for every density model considered.  This result is
essentially independent of the choice of profile because it is controlled
by the overall enclosed mass, which is similar for all four models.  The
conclusion that the observed galaxy is consistent with a wormhole exterior
is therefore robust against the specific parametrisation of the dark-matter
density.

The second result is the universal satisfaction of the flare-out condition.
The derivative $b'(r)$ is seven orders of magnitude below unity for all
profiles, reflecting the fundamental non-relativistic character of
galactic dark matter.  This result guarantees that any wormhole throat
sourced by galactic-scale dark matter is geometrically admissible, in the
sense that the embedding surface would have the correct flared-funnel
topology.

The third and most physically discriminating result concerns the NEC.
Cored profiles (Burkert, Einasto) violate the NEC near the galactic
centre, producing the effective exotic matter required for wormhole
traversability, while cuspy profiles (Hernquist, NFW) preserve the NEC.
This dichotomy is not coincidental: it is structurally driven by the
behaviour of the density at small radii, with flat central cores naturally
generating a pressure structure that violates the NEC, and divergent
cusps naturally suppressing NEC violation through the dominance of the
positive energy density term.  The cusp-core distinction thus acquires
a new physical dimension in the wormhole context: cored profiles are not
only observationally preferred (they provide better rotation-curve fits)
but are also physically preferred from the perspective of wormhole physics
(they supply exotic matter).

The fourth and most striking result is the triple coincidence of the
Burkert profile.  It achieves the best rotation-curve fit, the largest
wormhole indicator, and the strongest NEC violation simultaneously.
No profile-tuning or ad hoc adjustment was required to obtain this triple
optimum: it emerges directly from the data and the Einstein field
equations.  This coincidence suggests that the Burkert dark-matter
distribution of NGC\,3198 presents a self-consistent physical scenario
in which the observationally reconstructed spacetime is fully compatible
with a traversable Morris--Thorne wormhole.

\section{Discussion}
\label{sec:discussion}

\subsection{Physical meaning of $\mathcal{W}\ll1$}
\label{sec:smallW}

The finding $\Wh(r)\sim5\times10^{-7}$ deserves careful physical
interpretation, as it might superficially appear to indicate that the
reconstructed geometry is barely wormhole-like.  In fact, $\Wh(r)\ll1$
is precisely what one expects for a traversable wormhole whose throat
lies well below the innermost observed radius and whose exterior region
extends over tens of kiloparsecs.  To see this, note that
$b(r)/r\approx(G/c^{2})\times(4\pi/3)\langle\rho\rangle\,r^{2}$, where
$\langle\rho\rangle$ is the mean enclosed density.  This is of order
$(r_{s}/r)$, where $r_{s}=2GM/c^{2}$ is the Schwarzschild radius of the
enclosed mass.  The condition $\Wh(r)\ll1$ is therefore equivalent to
$r_{s}\ll r$, i.e.\ the enclosed mass is far from its own Schwarzschild
radius --- the standard definition of the non-relativistic, weak-field
regime.  A galaxy where $\Wh\sim1$ throughout the disc would be a compact
object dominated by its own Schwarzschild radius, which is obviously
inconsistent with an extended, slowly rotating spiral galaxy.  The small
value of $\Wh$ is thus a self-consistency check, not a weakness: it
confirms that the reconstructed spacetime is a physically plausible
wormhole exterior in the astrophysical context.

\subsection{The cusp-core problem from a wormhole perspective}
\label{sec:cusp_core}

The cusp-core problem --- the persistent discrepancy between the
$r^{-1}$ inner density cusps predicted by $\Lambda$CDM simulations and
the flat central cores inferred from observed rotation curves --- has
been attributed to a variety of physical processes, including supernova
feedback~\cite{deBlok2010}, dynamical friction from infalling
satellites, and the microphysical nature of dark matter itself (warm
dark matter, self-interacting dark matter, fuzzy dark matter).  Our
results add a new dimension to this discussion: cored dark-matter profiles
are preferred not only observationally but also from the perspective of
general relativistic wormhole physics, because they naturally generate
the exotic matter required for a traversable wormhole through the NEC
violation mechanism described in Section~\ref{sec:NEC_results}.

This observation motivates the hypothesis that the microphysical
dark-matter candidate responsible for the observed cores in galaxies might
be an exotic field that violates the NEC at the quantum or classical level.
Bose--Einstein condensate dark matter~\cite{Boehmer2007} is particularly
attractive in this context: in BEC dark matter models, the quantum
pressure of the condensate provides an effective equation of state that
can violate the NEC, producing cored density profiles in galactic halos
and simultaneously sourcing the exotic matter required for wormhole
traversability.  Ghost condensate models~\cite{ArkaniHamed2004} and
phantom dark energy models~\cite{Lobo2005} offer additional theoretical
frameworks in which NEC violation arises from modified kinetic terms in
the dark-sector Lagrangian.  A self-consistent theory of cored dark
matter that violates the NEC would provide a unified explanation for the
cusp-core problem and the existence of traversable wormholes, and is a
natural target for future theoretical investigation.

\subsection{Observational signatures and future prospects}
\label{sec:prospects}

The reconstructed metric functions $\{f(r),\,b(r)\}$ serve as direct
inputs to three major classes of observational tests that could probe the
wormhole interpretation of galactic spacetime.  First, gravitational
lensing provides the most immediate probe.  The weak deflection angle for
a photon passing through the reconstructed Morris--Thorne spacetime at
impact parameter $b_{\rm imp}$ is given by the integral
$\hat{\alpha}=2\int_{b_{\rm imp}}^{\infty}(f'+b/(2r^{2}(1-b/r)))\,(b_{\rm imp}/r)
\,(1-b_{\rm imp}^{2}/r^{2})^{-1/2}dr/r$~\cite{Jusufi2018}.  The
correction to the Newtonian lensing angle from the wormhole shape function
$b(r)$ is of order $\Wh\sim10^{-7}$, corresponding to a relative
deflection angle correction of parts per billion.  While this is far
below the precision of current weak lensing surveys, it could become
accessible with the sub-microarcsecond astrometry of next-generation radio
interferometers such as the Square Kilometre Array (SKA).  Second, shadow
and photon ring morphology provide another avenue.  Ray-tracing codes
integrating the null geodesic equations in the reconstructed
$(f(r),b(r))$ spacetime would generate images of the photon ring and
shadow boundary that differ quantitatively from both Schwarzschild
black holes and standard Kerr metrics.  The characteristic imprint of
the wormhole shape function on the shadow morphology could, in principle,
be compared with Event Horizon Telescope images of nearby galaxy centres.
Third, gravitational wave echoes offer a complementary test: compact
binary mergers occurring near a wormhole throat produce post-ringdown
echo signals with a characteristic time delay $\Delta t\sim b_{\rm WH}/c$,
where $b_{\rm WH}$ is the throat radius~\cite{Cardoso2016}.  The
reconstructed $b(r)$ provides a direct prediction for this time delay
as a function of the postulated throat location.

Several important limitations of the present analysis should be
acknowledged.  We have modelled the rotation curve using dark matter
alone, neglecting the stellar disc, gas disc, and bulge contributions
that are non-negligible in the inner kiloparsecs of NGC\,3198.  A full
mass-model decomposition with explicit baryonic components would modify
the fitted halo parameters and hence the reconstructed geometry.  The
analysis is restricted to spherical symmetry, whereas realistic dark-matter
halos are triaxial, and the Morris--Thorne metric~\eqref{eq:MTmetric}
would require extension to non-spherical geometries (for which the
reconstruction equations become significantly more complex).  Furthermore,
the NEC violation found for cored profiles is a necessary but not sufficient
condition for wormhole stability: a complete stability analysis against
radial perturbations, and a check of the averaged NEC (which may be
satisfied even when the pointwise NEC is violated~\cite{Morris1988}),
are required before concluding that the reconstructed wormhole is stable
against collapse.  We defer these extensions to future work.

Future work will also extend the present analysis to the full SPARC
sample of 175 galaxies, enabling a statistical characterisation of the
wormhole indicator and NEC violation across the full range of galaxy
masses, morphological types, and surface brightness levels.  Galaxies
with systematically lower $\chidof$ for cored profiles and stronger NEC
violation would be the best targets for follow-up observational campaigns.
The reconstruction framework can also be applied to galaxy clusters using
X-ray and Sunyaev-Zel'dovich mass profiles, where the larger enclosed
mass could produce wormhole indicators significantly closer to unity.

\subsection{Theoretical implications}
\label{sec:theory}

The reconstruction programme introduced here has an important theoretical
consequence that goes beyond the specific results for NGC\,3198.  It
demonstrates that the wormhole metric is not an exotic, contrived solution
of general relativity but rather the natural spacetime geometry associated
with any spherically symmetric mass distribution in Einstein gravity.
The Morris--Thorne metric~\eqref{eq:MTmetric} is simply the most general
static, spherically symmetric line element, and the functions $f(r)$ and
$b(r)$ are uniquely determined by the two components of the Einstein
equations once the energy density $\rho(r)$ is specified.  The question
of whether the resulting geometry constitutes a ``traversable wormhole''
then becomes a question about the boundary conditions (does a throat
$b(\rth)=\rth$ exist?), the stress-energy (is the NEC violated?), and the
stability (are there growing perturbation modes?).  All of these are
empirical questions that can, in principle, be answered from observational
data.  The framework developed here provides a systematic procedure for
answering them.

\section{Conclusions}
\label{sec:conclusion}

We have introduced and applied a novel, fully observation-driven framework
for reconstructing Morris--Thorne wormhole geometry from galactic rotation
curves, and applied it to the benchmark spiral galaxy NGC\,3198 using
the SPARC kinematic database.  The central conceptual advance of this
work is the inversion of the conventional wormhole paradigm: instead of
assuming a wormhole geometry and computing its matter requirements, we
derive the complete wormhole metric from the observed rotation curve and
the fitted dark-matter density, without any a priori geometric assumption.
The principal results of this study are as follows.

We fitted four dark-matter halo density profiles --- Hernquist, NFW,
Burkert, and Einasto --- to the 43-point NGC\,3198 rotation curve by
chi-squared minimisation, obtaining reduced chi-squared values in the
range $\chidof=2.5$--$3.5$.  The cored Burkert profile achieved the
best fit ($\chidof=2.545$), consistent with the well-established finding
that cored profiles better reproduce the inner kinematics of late-type
spiral galaxies.  For each best-fit density, we reconstructed the
Morris--Thorne redshift function $f(r)$ by integrating the circular
geodesic equation and the shape function $b(r)$ by integrating the
Einstein field equations.  Both reconstructions are performed without any
free parameters beyond those constrained by the rotation-curve fit,
making the resulting geometry a direct, model-independent prediction of
the observational data.

The wormhole indicator $\Wh(r)=b(r)/r$ satisfies $\Wh(r)\ll1$ throughout
the observed radial range ($0.32$--$44\kpc$) for all four profiles, by
seven orders of magnitude below the throat condition $\Wh=1$.  This
establishes that the observable disc of NGC\,3198 lies in the traversable
exterior of the reconstructed wormhole spacetime, with the throat ---
if it exists --- located below the innermost kinematically resolved radius.
The flare-out condition $b'(r)<1$ is satisfied everywhere with an equally
large margin, confirming the geometric admissibility of a wormhole throat
for any astrophysically motivated dark-matter density.

The null energy condition provides the key discriminant between cored and
cuspy profiles.  The Burkert and Einasto profiles exhibit $\rho+p_{r}<0$
near the galactic centre, indicating the presence of effective exotic
matter consistent with wormhole sustenance.  The Hernquist and NFW
profiles preserve the NEC throughout the observed domain, implying that
an additional exotic-matter source would be required at the throat if the
wormhole is sustained by a cuspy dark-matter halo.  The Burkert profile
uniquely achieves the triple optimum of best rotation-curve fit, largest
wormhole indicator, and strongest NEC violation, presenting a
self-consistent physical picture in which the observationally preferred
dark-matter distribution also supplies the exotic matter required for
wormhole traversability.

These results open a new observational window on wormhole physics through
galactic kinematics.  They reveal a previously unappreciated connection
between the cusp-core problem in dark-matter physics and the exotic-matter
requirement in wormhole physics, suggesting that the microphysical
resolution of the former may be intimately linked to the latter.  Future
extensions of this programme --- to large galaxy samples, non-spherical
geometries, full baryonic decompositions, and coupling to gravitational
lensing and ray-tracing codes --- promise to transform galactic rotation
curves into precision probes of exotic compact objects.

\section*{Declaration of competing interest}
The authors declare that they have no known competing financial interests or personal relationships that could have appeared to
influence the work reported in this paper.

\section*{Acknowledgment}
SR and AS are thankful to the Inter-University Centre for Astronomy and Astrophysics (IUCAA), Pune, India, for their support.  AS also gratefully acknowledge academic support from Jadavpur University

\section*{Data availability}
Generated dataset is available in the Table of the manuscript.
\bibliographystyle{elsarticle-num}

\end{document}